\documentclass[prb,showpacs,superscriptaddress,groupedaddress,floatfix,twocolumn]{revtex4-1}
\usepackage[latin1]{inputenc}
\usepackage[english]{babel}
\usepackage{graphicx}
\usepackage{amsmath}
\usepackage{amssymb}
\usepackage{amsbsy} 
\usepackage{amscd}
\usepackage{eucal}
\usepackage{color}
\usepackage{bm}
\usepackage[normalem]{ulem}
\usepackage{verbatim} 
\DeclareSymbolFont{mathbold}{OML}{cmm}{b}{it}
\DeclareMathSymbol{\bsigma}{\mathord}{mathbold}{27}

\begin{document}
\title{Signatures of spin-preserving symmetries in two-dimensional hole gases}
\author{Tobias Dollinger} 
\author{Michael Kammermeier} 
\author{Andreas Scholz} 
\author{Paul Wenk} 
\author{John Schliemann} 
\author{Klaus Richter} 
\affiliation{Institut f\"ur Theoretische Physik, Universit\"at Regensburg, D-93040 Regensburg, Germany}

\author{R. Winkler} 
\affiliation{Department of Physics, Northern Illinois University, DeKalb, Illinois 60115, USA}

\date{\today }
\begin{abstract}
  We investigate ramifications of the persistent spin helix symmetry
  in two-dimensional hole gases in the conductance of disordered
  mesoscopic systems. To this end we extend previous models by going
  beyond the axial approximation for III-V semiconductors. For
  heavy-hole subbands we identify an exact spin-preserving symmetry
  analogous to the electronic case by analyzing the crossover from
  weak anti-localization to weak localization and spin transmission as
  a function of extrinsic spin-orbit interaction strength.
\end{abstract}
\pacs{71.70.Ej,72.25.Dc,72.25.-b,72.15.Rn,73.63.Hs,73.63.-b}
\maketitle
\section{Introduction}
Control over spin relaxation is essential to the operational
capabilities of spin-based semiconductor
devices~\cite{fabian,schliemann-psh}. A major advance in this respect
has been the identification of an SU(2) symmetry that confines spin
evolution to a characteristic topology and allows realizations of
``persistent spin helix''~(PSH) excitations which are robust against
spin relaxation~\cite{schliemann-psh,bernevig}. The latter could be
identified by means of optical experiments~\cite{koralek, walser} in
two-dimensional electronic systems with linear-in-momentum
Bychkov-Rashba~\cite{rashba} and Dresselhaus~\cite{dresselhaus} type
spin-orbit interaction (SOI) (here depending on the parameters
$\alpha$ and $\beta$, respectively) of equal magnitude.
As had already been suggested in Refs.~\onlinecite{pikus-pikus, knap},
this symmetry at $\alpha=\pm\beta$ also becomes manifest in the weak
localization (WL) feature in magneto-conductance traces of disordered
materials with finite SOI, as opposed to weak antilocalization (WAL)
mediated by spin relaxation~\cite{hikami}. Recent experiments
confirmed theoretical predictions that the WL signature persists for
n-doped systems even in the presence of non-negligible intrinsic SOI
that scales with the cubic power of the wavenumber
$k$~\cite{kohda,golub-psh,luffe}.
The question naturally arises whether spin relaxation is also
suppressed in p-doped semiconductors. Therefore, in the present work
we investigate the generalization of the PSH symmetry arguments in the
context of structurally confined heavy-hole (HH) states in III-V
semiconductors forming a two-dimensional hole gas (2DHG).  In these
materials the spin is subject to strong SOI which typically enhance
spin relaxation.  This feature is mainly attributed to the carrier
density dependence of the spin splitting, that has been investigated
analytically by means of diagrammatic perturbation theory within the
spherical approximation for one or more
subbands~\cite{averkiev-golub1}.  Other works consider weak (anti-)
localization in hole gases based on a semianalytical~\cite{garate} as
well as a semiclassical and numerical~\cite{viktor-2dhg} treatment of
$4\times 4$ Luttinger-Kohn models~\cite{kohn-luttinger}.
\section{Two dimensional hole gas model}
Here we focus on strong confinement described by an effective $2\times
2$ model of the HH ground state. Our treatment is not restricted to
the spherical or axial approximations, which significantly widens the
range of observable phenomena compared to prior models.  The low
dimensionality allows for the identification of relevant symmetries
that are used to deduce optimum parameter regimes for controlling spin
relaxation. The structure of our model is given by the Hamiltonian
\begin{equation} 
{\rm H}={\rm H}_{\rm{kin}}\sigma_0+\bm{\Omega}_{\rm
2DHG}\cdot\bm{\sigma},\label{structure}
\end{equation}
where ${\rm H}_{\rm{kin}}$ denotes the kinetic energy, in which, to
first approximation, the explicit dependence on the Luttinger
parameters $\gamma_i$ enters in the effective mass,
$m_\text{eff}\approx m_0/(\gamma_1+\gamma_2)$, where $m_0$ is the mass
of the free electron and we have assumed a 2D system on a (001)
surface. \cite{winkler1} Here $\sigma_0$ is the identity matrix,
$\bm{\sigma}$ the vector of Pauli matrices, and $\bm{\Omega}_{\rm
  2DHG}$ the effective spin-orbit field coupling to the spin.  All
bold-face symbols used in the present text denote 2D vectors with only
$xy$ components. In contrast to the corresponding expression for
electrons, $\bm{\Omega}_{\rm 2DEG}$, where $k$-linear terms are
dominant~\cite{knap}, to leading order $\bm{\Omega}_{\rm 2DHG}$ is
characterized by a cubic momentum dependence. This is in agreement
with existing 2DHG models and results from coupling of the HH to the
light hole (LH) subbands~\cite{rashba-sherman, winkler2, bulaev-loss}.
The result~\eqref{structure} is obtained via a perturbation expansion
of the standard Luttinger-Kohn Hamiltonian~\cite{kohn-luttinger} in
the basis given by the subband edge states~\cite{winkler1}.  Our model
applies to typical zinc-blende structure materials, as can be inferred
from their material properties and calculated band structures given,
e.g., in Ref.~\onlinecite{winkler1}. In Eq.~\eqref{structure}, the
$2\times 2$ Hamiltonian $\rm H$ represents the subspace spanned by the
HH states of spin angular-momenta $\pm(3/2)\hbar$. The corresponding
hole spin-orbit field is given by
\begin{align}
&\bm{\Omega}_{\rm 2DHG}=\beta_{\rm HH}\bm{k} \label{HHH}\\ 
&+\lambda_{\rm{D}}\left\{-\bar{\gamma}\bm{k}^2\bm{k}+\delta[k_x^3\hat{\bm
      x}+k_y^3\hat{\bm y}-3k_xk_y(k_y\hat{\bm x}+k_x\hat{\bm
      y})]\right\}\nonumber\\ 
&+\lambda_{\rm{R}}\left\{\delta
  \bm{k}^2(k_y\hat{\bm x}+k_x\hat{\bm y})+\bar{\gamma}[-k_y^3\hat{\bm
      x}-k_x^3\hat{\bm y}+3k_x k_y \bm{k}]\right\}\nonumber
\end{align}
with the intrinsic Dresselhaus parameters
\begin{align} 
\beta_{\rm HH} ={}&
-\frac{\sqrt{3}C_k}{2}\label{linear},\\ \lambda_{\rm{D}} ={}&
\displaystyle \frac{ \sqrt{3}\hbar^2}{2 m_0 \Delta_{\rm HL}}\left[ C_k
  +\sqrt{3} \widetilde{b_{41}^{8v8v}} \left\langle
  k_z^2\right\rangle\right]\label{dcubic},
\end{align} 
the structural, electrical field $\left\langle E_z \right\rangle$ dependent
Bychkov-Rashba parameter, 
\begin{align} 
\lambda_{\rm{R}} ={}&\displaystyle \frac{3 \hbar^2}{2 m_0 \Delta_{\rm
    HL}}\left\langle E_z\right\rangle \widetilde{r_{41}^{8v8v}},
\end{align} 
and the Luttinger parameters $\bar{\gamma}=(\gamma_3+\gamma_2)/2$ and
$\delta=(\gamma_3-\gamma_2)/2$ as in
Ref.~\onlinecite{lipari}. Here $C_k$ is a material constant while
  $\widetilde{r_{41}^{8v8v}}$ and $\widetilde{b_{41}^{8v8v}}$ are
  parameters which depend on both material properties and geometry. In
  the bulk-case the latter two parameters coincide with
  $r_{41}^{8v8v}$ and $b_{41}^{8v8v}$ respectively, consistent with
  Ref.~\onlinecite{winkler1}. In the presence of a confinement,
  these parameters are modified. Since the value $b_{41}^{8v8v}$ is
  mainly defined by the valence band ($\Gamma_{8v}$) and
  conduction band ($\Gamma_{6c}$) gap $E_0$, this type of Dresselhaus
  contribution is hardly affected by the subband quantization. Thus,
  we assume $b_{41}^{8v8v}\approx\widetilde{b_{41}^{8v8v}}$. This does
  not hold for the dominant contribution by Rashba SOI, because the origin of the SOI, which is connected with the coefficient
  $\widetilde{r_{41}^{8v8v}}$, changes: In presence of the confinement,
  the contribution due to Rashba SOI in the effective HH system is
  dominated by the subband splitting between HH and LH. This dominant
  contribution is proportional to the term which is denoted here by
  $\widetilde{r_{41}^{8v8v}}$. The contribution described by
  $r_{41}^{8v8v}$, though, is induced by the coupling between valence and
  conduction bands: It represents a higher order correction and will not be
  considered in the following.\cite{tobepublished} Previous models~\cite{bulaev-loss,winkler2,bi}
  focus on the axially symmetric situation, $\delta=0$. The above expression represents a
  generalization of these models, allowing for the description of a broader range of materials and considering anisotropies that are important, for instance, in the
plasmon spectra of HH systems~\cite{andreas}.  Vertical confinement is
modeled by a potential well with perpendicular wavenumber
$\left<k_z^2\right>$ that displays a splitting $\Delta_{\rm
  HL}=2\gamma_2\hbar^2\left<k_z^2\right>/m_0$ between HH and
LH-bands. For further analysis, the terms proportional to the small
parameter $C_k$ are neglected in Eqs.~\eqref{linear} and
~\eqref{dcubic}, since for realistic materials and a narrow
confinement, the physics is dominated by the terms proportional to
$b_{41}^{8v8v} \left\langle k_z^2\right\rangle$, as shown in Table 6.3
in Ref.~\onlinecite{winkler1}.
Furthermore, the linear Dresselhaus term \eqref{linear} effectively
rescales the axially symmetric part of the cubic Dresselhaus
contribution.  Equation~\eqref{HHH} results from sequential
perturbation expansions up to third order in $k$ and to first order
with respect to the inverse splitting ${\Delta_{\rm HL}}^{-1}$ and to
$E_z$ imposed on the crystal. The identification of enhanced spin
relaxation times in this work is closely connected with broken axial
symmetry, since here a conserved quantity related to the spin degree
of freedom can only be constructed in the presence of terms with both
two- and threefold rotational symmetry in the extrinsic and the
intrinsic SOI\@.  Our findings suggest that obtaining an exact PSH
symmetry is limited by the parameters of realistic systems, since it
requires that $\gamma_3 = 0$. Although an approximate symmetry in the
leading-order Fourier components of $\Omega_{\rm 2DHG}$ causes a
weakly perturbed crossover from WAL to WL, similar to electronic
systems with cubic intrinsic SOI~\cite{kohda}, reaching exact
spin-preservation associated with $\gamma_3 = 0$ is not realistic.
This is due to the relation of the Kohn-Luttinger parameters described
in Ref.~\onlinecite{cardona}, which causes $\gamma_2$ to vanish in the
given case. This however violates the perturbation expansion, in which
the small parameter scales as $\gamma_2^{-1}$.
The influence of strain on our above discussed model has been
discussed in Ref.~\onlinecite{sacksteder}, in which the persistent
spin evolution requires the condition $\gamma_2=-\gamma_3$. This criterion is
not realizable for the above mentioned reasons.
In the present discussion we will also focus on the realization of
long-lived, but not fully preserved spin states in effective
heavy-hole models, for a suitable choice of the ratio
$\gamma_2/\gamma_3$. In contrast, for $\eta =-1$ and $\gamma_3=0$, as
investigated in Fig.~\ref{fig:2dhg1}, the fully symmetric situation is
obtained, corresponding to principally infinite spin-lifetimes.
\section{Charge transport analysis}
\subsection{Persistent spin helix conditions}
\begin{figure}
\centering \includegraphics[width=\columnwidth]{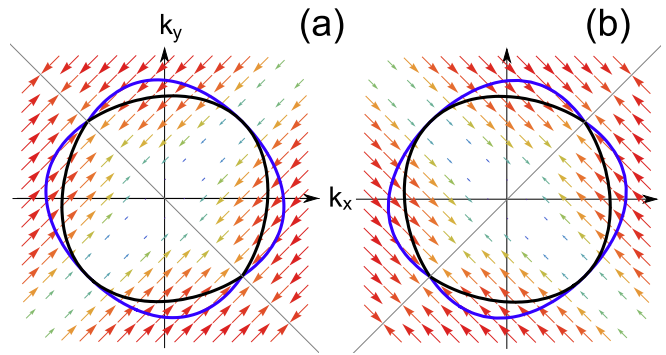}
\caption{\small{(Color online) Fermi surface for the different spin
    directions obtained from Eq.~\eqref{structure} (violet and blue
    contours) and the corresponding direction of the effective
    spin-orbit field, $\bf{\Omega}_{\rm 2DHG}$, illustrated by arrows.
    The SOI parameters establish a persistent spin helix for holes
    with uniaxial spin orientation corresponding to $\eta \equiv
    \lambda_{\rm R}/\lambda_{\rm D}=+1$ (a) and $\eta =-1$} (b).  In
  both cases the Luttinger parameters are $\bar{\gamma}=-\delta$,
  i.e., $\gamma_3=0$.  }
\label{fig:psh}
\end{figure}

The effect of the spin symmetry on the magneto-conductance $G(B)$ can be analyzed
by formulating the transmission in the Landauer-B\"uttiker
framework~\cite{landauer-1, buettiker},
\begin{equation}
 \frac{h}{e^2}G=\left(\sum_{n,m;\sigma=\sigma^{\prime}}
 +\sum_{n,m;\sigma\ne\sigma^{\prime}}\right)\left
 |t_{n\sigma,m\sigma^{\prime}}\right|^2 =: T_{\rm D}+T_{\rm
   OD},\label{spin-contributions}
\end{equation}
according to the spin quantum numbers $\sigma,\sigma'$ in terms of
diagonal spin-preserving channels $T_{\rm D}$ and spin off-diagonal
contribution $T_{\rm OD}$.  Here, $\sigma,\sigma' = \pm 1$ refer to an
arbitrary basis defined in the ballistic leads of a two-terminal
device representing our numerical model, while $n,m$ are integers that
define the transverse channel of the in- and outgoing states due to a
hard-wall confinement defining the edges of the leads.  The lead
wavefunctions $\left|\phi_{n,\sigma}\right>$ and
$\left|\phi_{m,\sigma^{\prime}}\right>$ enter into the Fisher-Lee
relation for the amplitudes
\begin{align}
t_{n\sigma,m\sigma^{\prime}}\propto\int_{\partial\text{Leads}}{\rm
  d}^2r\left<\phi_{n,\sigma}\right|\left.y_1\right>\left<y_1\right|G_{\rm
  R}\left|y_2\right>\left<y_2\right.\left|\phi_{m,\sigma^{\prime}}\right>,
\end{align}
where the integration is taken over the lead cross sections~\cite{fisher-lee}.
$G_{\rm R}=\left(E_{\rm F}-{\rm H}+0_{+}\right)^{-1}$ is the Green's function of
the scattering region at fixed Fermi energy $E_{\rm F}$.
Knap et al.~\cite{knap} found in n-type systems particular relations
between extrinsic and intrinsic SOI magnitude, for which the Cooperon
becomes separable and a WL signal rather than WAL is observed. In
terms of the structure provided by Eq.~\eqref{spin-contributions},
$T_{\rm OD}$ vanishes in this case and correspondingly spin scattering
is absent even in transport in disordered systems.  This is equivalent
to the observation that the system displays an exact, disorder
independent symmetry~\cite{schliemann-psh,bernevig}, which allows for
a decomposition within the corresponding constant eigenbasis
$\left\{\left|\chi_{\sigma}\right>\right\}$ into
$\bm{\Omega}\cdot\bm{\sigma}=\sum_{\sigma=\pm1}E_{\sigma}(\bm{\Omega})\left|\chi_{\sigma}\right>\left<\chi_{\sigma}\right|$.
Hence, when taking the spin trace in Eq.~\eqref{spin-contributions} in
the basis $\left\{|\chi_{\sigma}\rangle=(1,\sigma\exp[\pm
  i\pi/4])^\dagger\right\}$, corresponding to the existence of the
conserved quantity $\Sigma_{\pm}=\sigma_x\pm\sigma_y$ or,
equivalently, fixed in-plane spin orientation along
$\varphi=\pm\pi/4$, one finds that
\begin{align}
T_{\rm
  OD}~\propto~\sum_{\sigma\neq\sigma^{\prime}}\left|\left<\chi_{\sigma}\middle|\chi_{\sigma^{\prime}}\right>\right|^2=
\sum_{\sigma\neq\sigma^{\prime}}\delta_{\sigma,\sigma^{\prime}}
\end{align}
is suppressed and $T_{\rm D}$ decomposes into two independent channels
which trivially display WL~\cite{knap}.
In the hole model~\eqref{HHH} we find the analogue to the electronic PSH
symmetry if the system parameters fulfill $\eta =\pm1$
and  $\bar{\gamma}=-\delta$, i.e., $\gamma_3=0$, where we define the parameter
$\eta \equiv \lambda_{\rm R}/\lambda_{\rm D}$.  In these two cases the
direction of $\bm{\Omega}_{\rm 2DHG}$ is fixed independently of the momentum, more
precisely by 
\begin{align}
\bm{\Omega}_{\rm 2DHG}\propto{}&[-{\bm k}^2(k_x\pm k_y)\pm 3k_x
k_y(k_x\pm k_y)\nonumber\\
{}&-k_x^3\mp k_y^3](\hat{\bm x}\pm\hat{\bm y}).  
\end{align}
We illustrate these cases in Fig.~\ref{fig:psh}, where the effective
spin-orbit field ${\bm \Omega}_{\rm 2DHG}$ is oriented along a fixed
direction for both spin split subbands. The structure of
Eq.~\eqref{HHH} implies an additional symmetry for
$\bar{\gamma}=\delta$, which is however outside the range of validity,
since it corresponds to $\gamma_2=0$ and thereby a breakdown of
perturbation theory.  Although the given parameters can be engineered
in realistic material systems, as indicated by Table C.9 in
Ref.~\onlinecite{winkler1}, it is not possible to influence the
effective values of $\gamma_3$, without simultaneously changing
$\gamma_2$ or the effective values of the Rashba and Dresselhaus
coefficients~\cite{cardona}.

\subsection{Numerical Setup}
To investigate the previously described properties we simulate
transport in disordered hole systems connected to two terminals,
represented by ballistic semi-infinite leads without SOI\@. The latter
is switched on and off adiabatically over one fifth of the total
length of a rectangular scattering region to which the leads are
connected.  We use an average over an Anderson-like uniformly
distributed random-box potential $V_{\rm{dis}}$ to simulate disorder.
The perpendicular magnetic field is included by means of Peierls'
substitution.  The Hamiltonian is then discretized on a tight-binding
grid in position space and the transmission amplitudes are obtained by
an optimized recursive Green's function algorithm~\cite{mike}.  Since
we are interested in modeling bulk transport, we implemented periodic
boundary conditions in the transverse direction to minimize effects
from the boundaries.

\subsection{PSH signatures in the magneto-conductance}
\begin{figure}
\centering
\includegraphics[width=1.0\columnwidth]{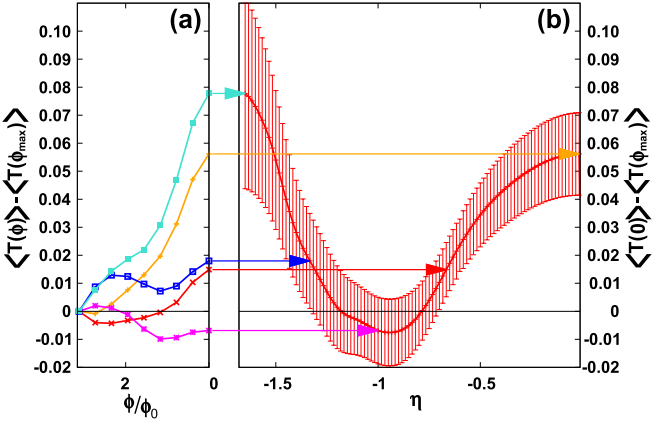}
\caption{\small{(Color online)} Signatures of spin-preserving
  symmetries in weak localization of a two-dimensional hole gas. (a)
  Disorder-averaged magneto conductance correction $\left\langle
  T(\phi)\right\rangle-\left\langle T(\phi_{\rm max})\right\rangle$ as
  a function of flux $\phi$ (in units of $\phi_0=h/e$; $\phi_{\rm
    max}/\phi_0=3.1$) for spin-orbit coupling ratios $\eta \equiv
  \lambda_{\rm R}/\lambda_{\rm D}=-1.65, -0.017, -1.33, -0.67, -1 $
  (from top to bottom).\\ (b) Conductance correction $\left\langle
  T(0)\right\rangle-\left\langle T(\phi_{\rm max})\right\rangle$ as a
  function of $\eta$. Negative magneto conductance reflects
  suppression of spin relaxation close to $\eta=-1$. System
  parameters: Disorder average over 1000 impurity configurations for a
  scattering region of aspect ratio (length $L$ to width $W$) 200:80
  unit cells with periodic boundary conditions in transverse
  direction.  Quantum transmission computed for $k_{\rm F}W/\pi=13$
  hole states per spin supported in the leads, elastic mean free path
  $l=0.04~W$, $\gamma_3=0$, and fixed Dresselhaus spin precession
  length $k_{\rm D} W\approx 1$, defined below Eq.~\eqref{Q_2dhg}.  }
\label{fig:2dhg1}
\end{figure}
The symmetry condition becomes apparent in the magneto-conductance of
disordered 2DHG systems, as illustrated in Fig.~\ref{fig:2dhg1}.  Here
we show results of the numerically calculated disorder averaged
transmission, Eq.~\eqref{spin-contributions}, for finite cubic
intrinsic SOI $\lambda_{\rm D}$ as a function of the extrinsic SOI
$\lambda_{\rm R}$.  For demonstrative purposes, we set $\gamma_3=0$.
Representative examples of the conductance correction traces are shown
in Fig.~\ref{fig:2dhg1}~(a), which display typical WAL and WL
lineshapes as a function of magnetic flux $\phi$ from a homogeneous
magnetic field perpendicular to the 2DHG plane.  Considering the
dependence on $\eta$, we find pronounced signatures of WAL if $\eta$
is far from $-1$.  When $\eta$ approaches $-1$, a crossover from WAL
to WL occurs as indicated by a maximum negative conductance
correction, in agreement with the symmetry argument. In
Fig.~\ref{fig:2dhg1}~(b) our results are summarized in terms of the
conductance at maximum magnetic flux $\left\langle T(\phi_{\rm
  max})\right\rangle$ subtracted from the correction at zero flux
$\left\langle T(0)\right\rangle$ plotted as a function of $\eta$,
where we chose $\phi_{\rm max}=3.1~\phi_0$.  The results show that the
parameter regime where a PSH type symmetry occurs is characterized by
a negative conductance correction, i.e., by a WL signature.

\subsection{Diagrammatic Approach}
\subsubsection{Cooperon of the Effective 2DHG}
Our analysis above is confirmed within a diagrammatic perturbative
treatment by exact diagonalization of the Cooperon $\hat C({\bf Q})$
in the framework of the effective model~\eqref{structure}. For this
purpose, the scheme presented in Refs.~\onlinecite{kettemann,paul} for
electrons is generalized to holes. The diagrammatic approach is
justified since we assume the system to be in the diffusive regime
fulfilling the Ioffe-Regel criterion, $E_{\rm F}\tau/\hbar\gg 1$, with
elastic scattering time $\tau$ and Fermi energy $E_{\rm F}$. Here the
scattering is modeled by standard ``white-noise'' disorder $V({\bf
  x})$ which vanishes on average, $\langle V({\bf x}) \rangle=0$, and
is uncorrelated, $\langle V({\bf x})V({\bf x^\prime}) \rangle=
\delta({\bf x}-{\bf x^\prime})/(2\pi\nu\tau/\hbar)$, where $\nu$ is
the density of states per spin channel. Unfortunately, a general
analytical study of the Cooperon of the confined 2DHG including both
the SOI due to BIA and SIA is spoiled by the fact that this operator,
which is necessary to describe the conductivity correction due to
interference between a $J=3/2$ hole and its time-reversed counterpart,
requires 16 dimensions. However, since we mapped the $4\times 4$
Luttinger Hamiltonian onto an effective one considering only the spin
$\pm 3/2$ subspace, the Cooperon of this effective model is equivalent
to the Cooperon of s-band conduction electrons presented in
Refs.~\onlinecite{kettemann,paul}, except for the absolute value of
the spin and the terms appearing due to SOI\@. The total Cooperon
momentum ${\bf Q}$ is the sum of the momenta of the retarded and
advanced propagators of holes, ${\bm Q}={\bm k}+{\bm
  k}^{\prime}$. Their spins $(3/2)\hbar{\bm\sigma}$ and
$(3/2)\hbar{\bm\sigma}^{\prime}$ sum up to ${\bm S}=(3/2)\hbar({\bm
  \sigma}+{\bm \sigma}^{\prime})$. We get to second order in
$(\hbar{\bf Q}+(2/3)m_\text{eff}\hat a {\bf S})$ and after an angular
average $\langle\ldots\rangle_\varphi$ over the Fermi surface
\begin{align}
	   \hat C({\bf Q})=\frac{\hbar}{D_h\left(\hbar{\bf Q}
	   +\frac{2}{3}m_\text{eff}\left\langle\hat
	   a\right\rangle_\varphi\cdot{\bf S}\right)^2+H_\gamma},
	   \label{cooperon}
\end{align}
where $D_h=\tau v_{\rm F}^2/2$ is the diffusion constant. The matrix
$\hat{a}$ in the effective vector potential term is defined by the
relation ${\bm \sigma}\cdot\bm{\Omega}_{\rm{2DHG}}={\bm k}\cdot(\hat{
  a}\cdot{\bm \sigma)}$. With $\langle\hat
a\rangle_\varphi\equiv\hat\alpha$ we find
\begin{align}
 \hat\alpha={}&\frac{m_\text{eff}}{2\hbar^4}E_{\rm F}
 \left(\begin{array}{cc} \lambda_{\rm D}(3+c_{\rm D}) & \lambda_{\rm
     R}(3+c_{\rm R})\\ \lambda_{\rm R}(3+c_{\rm R}) & \lambda_{\rm
     D}(3+c_{\rm D})
\end{array}\right),
\end{align}
with $E_{\rm F}=m_\text{eff}v_F^2/2$ the Fermi energy, $c_{\rm
  D}=2\gamma_3/\gamma_2-1$, $c_{\rm R}+c_{\rm D}=-2$. The term
\begin{align}
H_\gamma={}&\frac{1}{9}\frac{D_h m^4_\text{eff}E_{\rm
    F}^2}{\hbar^8}\left[[(\lambda_{\rm D}^2(c_{\rm
      D}-1)^2+\lambda_{\rm R}^2(c_{\rm
      R}-1)^2)\right.\nonumber\\ {}&\left.(S_x^2+S_y^2)+2(c_{\rm
      D}-1)(c_{\rm R}-1)\lambda_{\rm R}\lambda_{\rm
      D}\{S_x,S_y\}\right]
\end{align}
is ${\bm Q}$-independent and resembles the corresponding expression
for the 2DEG in Ref.~\onlinecite{kettemann} which appears due the
existence of cubic Dresselhaus SOI\@. We simplify the calculation by
assuming $\beta/\lambda_{\rm D}$ to be negligibly small and by
rescaling the Cooperon Hamiltonian $H_\text{C}\equiv {\hat C}^{-1}$
for nonzero intrinsic Dresselhaus $\lambda_{\rm D}$:
\begin{align}
\tilde H_\text{C}\equiv{}& \frac{H_\text{C}}{D_h\left(\lambda_{\rm D}\frac{m^2_\text{eff}}{3 \hbar^3}E_{\rm F}\right)^2}\\
={}&
\left(\begin{array}{c}
{\hbar\tilde Q_x} +\hbar^{-1}\left[(3+c_{\rm D})S_x+\eta(1-c_{\rm D}) S_y\right] \\
{\hbar\tilde Q_y} +\hbar^{-1}\left[(3+c_{\rm D})S_y+\eta (1-c_{\rm D}) S_x\right]
\end{array}\right)^2\nonumber\\
{}&+\hbar^{-2}[(1-c_{\rm D})^2+\eta^2 (3+c_{\rm D})^2](S_x^2+S_y^2)\nonumber\\
{}&+2\hbar^{-2}(1-c_{\rm D})(3+c_{\rm D})\eta\{S_x,S_y\},
\end{align}
with $\tilde Q_i=Q_i/\lambda_{\rm D}\frac{m^2_\text{eff}}{3
  \hbar^3}E_{\rm F}$. Since the spectra of the Cooperon and Diffuson
are equal as long as time reversal symmetry is not broken, the term
$H_\gamma$, which cannot be rewritten as a vector potential, causes in
general gaps in the triplet-sector of the spectrum which correspond to
finite spin relaxations~\cite{paul}. As in the case of 2DEG, only the
triplet sector is affected by SOI (here, due to the effective HH
model, we have $S=(3/2+3/2)\hbar$ but one can use $\{|S=0,m=0\rangle$,
$|S=1,m=1\rangle$, $|S=1,m=0\rangle$, $|S=1,m=-1\rangle\}$ as a basis
since ${\bf S}\sim(\bsigma+\bsigma^\prime)$ ).  The appearance of
gapless modes besides the singlet mode, i.e., the existence of
persistent spin states as found by using the Landauer-B\"uttiker
framework, will be discussed in the following.
\subsubsection{Persistent and Long-Lived Spin States}
We focus on two interesting parameter regimes: Luttinger parameters
which describe systems close to axial symmetry where we have $c_{\rm
  D}\approx 1$ and the extreme case $c_{\rm D}=-1$ for which the SO
field $\Omega_\text{2DHG}({\bf k})$ is aligned in one direction if
$|\eta|=1$ as presented in Fig.~\ref{fig:2dhg1}.

An analysis of the Cooperon triplet spectrum for values $c_{\rm
  D}\approx 1$ and moderate strength of Rashba SOI, i.e., $-\sqrt{3}
\le \eta \le \sqrt{3}$, reveals that the absolute minimum expressed in
polar coordinates as $\tilde{\bm{Q}} = (\tilde Q,\varphi)$ can be
found at finite $\tilde Q_\text{min}=\sqrt{3(3-\eta^2)(5+\eta^2)}$
with an energy of
\begin{align}
{}&\tilde E_\text{min,1}(\tilde Q_\text{min})= 21+66\eta^2-3\eta^4\nonumber\\
{}&+\frac{3}{2}\lambda
[\eta(5+6\eta^2+\eta^4)\sin(2\varphi)-7-22\eta^2+\eta^4]+\mathcal{O}(\lambda^2),
\end{align}
with $\lambda=1-c_{\rm D}$, $|\lambda|\ll 1$. Thus, the spectrum will
always be gapped with a minimal gap for $\eta=0$. The spin states to
which the minima correspond are long-lived (finite spin relaxation)
modes which describe a spin helix due to $\tilde
Q_\text{min}>0$.\cite{paul} The situation differs completely for the
case where the Luttinger parameter $\gamma_3$ vanishes, i.e., $c_{\rm
  D}=-1$. We find an absolute minimum of the Cooperon triplet spectrum
at $\tilde Q=0$ with
\begin{align}\label{Eq:Emin-1}
\tilde E_\text{min,-1}(\tilde Q)={}& 24(1-|\eta|)^2+\frac{1}{4}\tilde
Q^2[(3-|\eta|)(1+|\eta|)\nonumber\\ {}&
  +(1-|\eta|)^2\sin(2\varphi)]+\mathcal{O}(\tilde Q^3),
\end{align}
for $|\eta|\approx 1$. As a consequence, we obtain a gapless mode for $|\eta|=1$.
This supports the numerical findings of persistent spin states if the
aforementioned symmetries are present: Changing $\eta$ from $\eta=0$
to $\eta=-1$ as done in Fig.~\ref{fig:2dhg1}, we see that the
energetically lowest mode, Eq.~\eqref{Eq:Emin-1}, is gapped at
$\eta=0$. Thus, the negative contribution of the triplet modes to the
conductivity correction $\Delta\sigma=(W/L)(e^2/h) \langle
T(0)\rangle-\langle T(\phi_\text{max})\rangle$ is suppressed and we
end up with an enhancement of conductivity (WAL) stemming from the
positive gapless singlet channel. Enhancing $|\eta|$ does not change
the singlet-mode contribution to $\Delta\sigma$. However, the
suppression of triplet-contribution is reduced: We see a reduction of
conductivity leading to WL in the case where in addition to the
gapless singlet mode a gapless triplet mode appears.
\begin{figure}
 \centering
  \includegraphics[width=1.0\columnwidth, angle=-90]{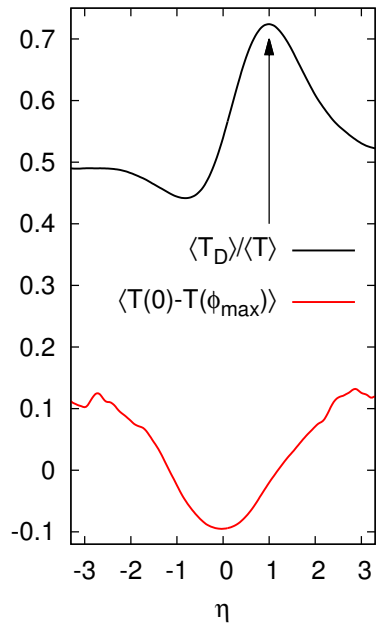}
\caption{Top: Ratio of disorder averaged diagonal transmission over
  total transmission $\left\langle T_{\rm D}\right\rangle/\left\langle
  T\right\rangle = \left\langle T_{\rm D}\right\rangle/\left\langle
  T_{\rm D}+T_{\rm OD}\right\rangle$ as a function of $\eta$ for a
  scattering region with 150:80 aspect ratio for fixed Dresselhaus
  spin precession length $k_{\rm D}^{-1}\approx (k_F^2\lambda_D)^{-1}=1.3~W$ and Luttinger parameters
  $\gamma_2=1$ and $\gamma_3=0.25$ for which no exact spin-preserving
  symmetry can be established.  The peak of $\left\langle T_{\rm
    D}\right\rangle/\left\langle T\right\rangle$ at $\eta=1$
  (indicated by the arrow), coincides with the maximum of the diabatic
  transition probability of Eq.~\eqref{Q_2dhg}.  Average Transmission
  shown includes 1000 disorder configurations. With respect to the
  eigenbasis of $\eta=-1$, we obtain a curve that coincides with the
  mirror image of the shown plot, displaying a maximum at $\eta=-1$.
  Bottom: amplitude of the magneto-conductance correction in which no
  WAL-WL-WAL transition is observable in the vicinity of the
  symmetry point of $|\eta|=1$ due to insufficient spin randomization
  in the regime $|\eta|<1.19$.  This example indicates that in a sweep
  of the Rashba SOI, the point where Rashba and Dresselhaus SOI are
  balanced displays a clear signal in the diagonal transmission, while
  it may not be detectable in form of a WAL-WL-WAL transition.  }
\label{fig:2dhg2}
\end{figure}
\section{Spin transport analysis}
Apart from considering the indirect influence of the PSH symmetry on
the WL-WAL transition, it seems natural to search for a manifestation
of a symmetry in $T_{\rm D}$, Eq.~\eqref{spin-contributions}, since
its effects could be determined by magnetic polarization of the leads,
allowing for spin transistor operation even in the presence of
disorder~\cite{schliemann-psh}.  Numerically we can confirm the
validity of the latter approach by calculating the normalized quantity
$T_{\rm D}/(T_{\rm D}+T_{\rm OD})$ as a function of $\eta=\lambda_{\rm
  R}/\lambda_{\rm D}$, as shown in Fig.~\ref{fig:2dhg2}. We identify a
pronounced transmission maximum at $\eta=1$ in the basis corresponding
to the $+\pi/4$ spin orientation even in situations where the exact
PSH-type symmetry is not realized. In the given example we chose the
Luttinger parameters $\gamma_2=1$ and $\gamma_3=0.25$ which correspond
to arbitrarily chosen parameters, that serve as a proof of concept of
measurements in a setup, where a WAL-WL-WAL transition upon variation
of the Rashba SOI is not experimentally observable.  For values of
$|\eta|<1.19$, the conductance correction corresponding to
Fig.\ref{fig:2dhg2} is still dominated by WL,
corresponding to insufficient spin randomization. For
$|\eta|>1.19$ we observe WAL, consistent with the increased magnitude
of the SOI. Therefore, the experimental determination of the relative
magnitude of Rashba and Dresselhaus SOI from the WAL-WL-WAL transition
is not feasible in this setup, because the point where both
contributions are in balance, i.e. $|\eta|=1$, lies within the regime
of small spin randomization.  In the spin-resolved transmission signal
the symmetry point is however clearly visible, as demonstrated by the
maximum in Fig.~\ref{fig:2dhg2}.  For parameters far from $\eta=1$ the
spin transmission is equally distributed among the diagonal and
off-diagonal channels.
When $|\eta|$ approaches unity, $T_{\rm D}$ formally corresponds to
the probability of diabatic Landau-Zener transitions between
instantaneous eigenstates $\left|\pm\bm{\Omega}\right\rangle = (1,\pm
\exp[-i\arctan\left(\Omega_y/\Omega_x\right)])^\dagger$ of the
spin-orbit contribution~\eqref{HHH}.  The momentum direction is
changed by disorder scattering such that the spin evolution is subject
to inhomogeneities of the effective spin-orbit field $\bm{\Omega}$.
At the minima of the anisotropic spin splitting $2
\left|E_{\sigma}(\bm{\Omega})\right|$, this induces transitions of the
type $\left|\pm\bm{\Omega}\right>\to\left|\mp\bm{\Omega}\right>$ with
Landau-Zener~\cite{landau,zener-transition} transition probability
\begin{equation}
P_{\rm{D}}=\exp\left[-2\pi{\epsilon_{12}^2}/\left({\hbar\left|\partial_t\left(\epsilon_1(t)-\epsilon_2(t)\right)\right|}\right)\right].\label{lz-template}
\end{equation}

The value of $P_{\rm{D}}$ is calculated from the minimal spin splitting, $2
\epsilon_{12}$, in the corresponding directions $\varphi:=\arctan(k_y/k_x)$ and
the slope of the splitting, $\epsilon_1(t)-\epsilon_2(t)$, between the fully
diabatically coupled basis states.

These transitions enhance the value of $T_{\rm D}$ while completely
suppressing $T_{\rm OD}$ for $P_{\rm{D}}=1$.  The spinors
$\left\{\left|\chi_{\sigma}\right>\right\}$ underlying
Eq.~\eqref{spin-contributions} coincide with the diabatic
superposition of the states $\left|\pm\bm{\Omega}\right>$.  The latter
can be checked by considering
$\left<\chi_{\sigma}\right|\bm{\Omega}\cdot\bm{\sigma}\left|\chi_{\sigma}\right>$.
Within the HH model~\eqref{HHH} the diabatic basis coincides with that
of the PSH eigenstates $\left\{\left|\chi_{\sigma}\right>\right\}$ of
a 2DEG~\cite{schliemann-psh}.  For p-type systems we find a
probability of
\begin{equation}
\ln(P_{\rm D})_{\rm{2DHG}}=-\zeta l |k_{\rm
  D}|\left|\bar{\gamma}+\delta\right|\left(1-|\eta|\right)^2,\label{Q_2dhg}
\end{equation}
with the elastic mean free path $l$ and a phenomenological factor
$\zeta$ of order 1, related to the details of the scattering. These
quantities enter together with the transport time $\tau$ into the rate
of change in angle $\varphi$ in the relation $\delta\varphi=\pi \delta
t/(2 \tau \zeta)$.  The characteristic lengthscale of spin precession
$k_{\rm D}^{-1}$ is approximated as $k_{\rm D}\approx k_{\rm
  F}^2\lambda_{\rm D}$.  Equation~\eqref{Q_2dhg} is derived under the
assumption that $\bar{\gamma}\neq-\delta$. Although the
expression~\eqref{lz-template} for the Landau-Zener transition
probability predicts a clear maximum at $|\eta|=1$, Eq.~\eqref{Q_2dhg}
does not cover the description of $T_{\rm D}$ for parameters where the
PSH symmetry is established. It is nevertheless applicable to
realistic material parameters if $\gamma_3\neq0$ and, consequently,
$\bar{\gamma}\neq-\delta$, which is verified by a numerical transport
analysis.
The analysis of $T_{\rm D}$ can be applied to electronic systems as
well, with an effective spin-orbit field,
\begin{equation}
\bm{\Omega}_{\rm{2DEG}} =\alpha\bm{k}\times\hat{\bm{z}}+\beta\left(k_x \hat{\bm{x}}- k_y\hat{\bm{y}}\right)
+\gamma \left(-k_x k_y^2\hat{\bm{x}}+k_y k_x^2\hat{\bm{y}}\right),\label{beff2DEG}
\end{equation}
for transport along the [100] direction in a 2DEG grown in [001]
direction and with $\left<k_z^2\right> \gamma=\beta$.\cite{knap} In
systems described by this model the corresponding Landau-Zener
transition probability is given by
\begin{equation}
\ln(P_{\rm{D}})_{\rm{2DEG}}=-\zeta l|k_{\beta}|\left(\Gamma_{\beta}/2-1\pm\eta\right)^2,\label{Q_2deg}
\end{equation}
with the Dresselhaus spin precession length
$k_{\beta}^{-1}=\left(m_{\rm{eff}}\beta/\hbar^{2}\right)^{-1}$, ratio
of cubic and linear SOI $\Gamma_{\beta}=k_{\rm F}^2\gamma/\beta$ and
the phenomenological factor $\zeta$ as it appears in
Eq.~\eqref{Q_2dhg}. This model has been verified by numerical
calculations which are beyond the scope of this work.
In both p- and n-type systems, the signatures in $T_{\rm D}$ are
robust against disorder.

Therefore, as an experimental approach to analyzing spin relaxation
lengths in transport within HH systems, a detection of the PSH
signature in the longitudinal conductance of a spin-polarized current
is favorable.  The mechanism responsible for the peaks in $T_{\rm D}$
is the momentum space analogue to the effect of a spatially
inhomogeneous helix-type Zeeman term on the spin conductance in dilute
magnetic semiconductors~\cite{betthausen}.  An alternative measurement
method for further investigation of the HH PSH is represented by
magneto-optical Kerr rotation techniques, which recently allowed to
map the spin topology in electronic systems~\cite{walser}.
\section{Acknowledgements}
We acknowledge financial support by DFG within the collaborative
research center SFB 689 and by the Elitenetzwerk Bayern (T.D.).  We
thank J. Fischer and V.  Kr\"uckl for helpful discussions, and
M. Wimmer for providing the numerical algorithm used here.
\appendix
\section{Diabatic Transitions in Momentum Space}
To obtain the leading-order contribution to the spin-diagonal
transmission defined in Eqs.~\eqref{Q_2dhg} and~\eqref{Q_2deg}, we
start from the diagonal approximation to the semiclassical
transmission amplitudes~\cite{frustaglia},
\begin{equation}
 T_{\sigma,\sigma} \sim \sum_{\gamma} \left|A_{\gamma}\right|^2\left|\left\langle\sigma\right|{\rm D}_{\gamma}\left|\sigma\right\rangle\right|^2, \label{semiclassical-transmission}
\end{equation}
with the stability amplitude $\left|A_{\gamma}\right|^2$ corresponding
to classical paths $\gamma$ that connect the incident lead with the
outgoing lead for the respective channels.  Summing above expression
with respect to the spin polarizations $\sigma = \pm1$ yields the
semiclassical leading order contribution to $T_{\rm D}$ after
performing a disorder average.  Without the spin evolution kernel $\rm
D$, the Drude conductance can be estimated from
Eq.~\eqref{semiclassical-transmission}, since the summation over the
amplitudes can be expressed in terms of classical transmission
probabilities~\cite{chakravarty-schmid}.
For small spin splitting compared to the kinetic energy, the
trajectories $\gamma$ are solely determined by classical properties of
the system. They parameterize the spin dynamics via the equation for
the spin evolution kernel along the path $\gamma$~\cite{frustaglia},
\begin{equation}
 i\hbar \frac{\partial}{\partial t}{\rm D}_{\gamma}(t)\left|\sigma\right\rangle
 = {\bm \Omega}({\bm k}(t))\cdot {\bm \sigma}{\rm D}_{\gamma}(t)
 \left|\sigma\right\rangle,  \label{spin-kernel}
\end{equation}
for the effective spin-orbit field $\bm \Omega$ for electrons,
Eq.~\eqref{beff2DEG}, or holes, Eq.~\eqref{HHH}, respectively.
To estimate for which values of the spin-orbit parameters the value of
$T_{\sigma, \sigma}$ reaches a maximum, we calculate
$\left|\left\langle\sigma\right|
D_{\gamma}(t)\left|\sigma\right\rangle\right|^2$ from
Eq.~\eqref{spin-kernel} as the probability to remain in the
instantaneous eigenstate matching the initial polarization at the
lead-cavity interface via the Landau-Zener
formula~\cite{landau,zener-transition}.  We specified the
time-dependent problem~\eqref{spin-kernel} after disorder average by a
momentum ${\bm k }(t)\approx k(\cos\varphi(t) \hat{\bm
  x}+\sin\varphi(t) \hat{\bm y})$ that changes due to elastic
small-angle scattering according to $\delta\varphi=\pi \delta t/(2
\tau \zeta)$. Here we introduce the phenomenological parameter $\zeta$
by hand. $\zeta = 1$ corresponds to a momentum change due to isotropic
scattering and $\tau$ is the elastic momentum relaxation time.  Note
that the disorder model on which the numerical results of
Fig.~\ref{fig:2dhg2} are based, consists of Anderson-like impurity
configurations with small correlation lengths.  Although the
semiclassical picture presented above is not applicable to this setup
in a strict sense, it describes the observed behavior remarkably well.

\end{document}